# Dark mammoth trunks in the merging galaxy NGC 1316 and a mechanism of cosmic double helices

Per Carlqvist

*Space and Plasma Physics, School of Electrical Engineering, Royal Institute of Technology, SE-100 44 Stockholm, Sweden*

**Abstract.** NGC 1316 is a giant, elliptical galaxy containing a complex network of dark, dust features. The morphology of these features has been examined in some detail using a Hubble Space Telescope, *Advanced Camera for Surveys* image. It is found that most of the features are constituted of long filaments. There also exist a great number of dark structures protruding inwards from the filaments. Many of these structures are strikingly similar to elephant trunks in H II regions in the Milky Way Galaxy, although much larger. The structures, termed *mammoth trunks*, generally are filamentary and often have shapes resembling the letters V or Y. In some of the mammoth trunks the stem of the Y can be resolved into two or more filaments, many of which showing signs of being intertwined. A model of the mammoth trunks, related to a recent theory of elephant trunks, is proposed. Based on magnetized filaments, the model is capable of giving an account of the various shapes of the mammoth trunks observed, including the twined structures.

## 1 Introduction

Most elliptical galaxies are known to be poor in dust and cold gas. Nevertheless, there exist exceptions to this rule comprising ellipticals exhibiting pronounced dust features. A striking example is the giant, elliptical galaxy NGC 1316 showing an intriguing network of dark dust lanes. Basically, NGC 1316 is a Morgan D-type galaxy with an unusually small and bright core. Almost from the very center of the galaxy out to radial distances of more than one minute of arc, a complex system of dark dust features stands out in silhouette against the bright background of unresolved stars. The presence of non-concentric shells, loops, and tails in the outer parts of the galaxy, together with weak ripples in the inner regions, strongly suggests that NGC 1316 is a merger remnant (Schweizer, 1980; Bosma, Smith & Wellington, 1985). One or more gas-rich, disk galaxies are thought to have collided with the progenitor of NGC 1316 and merged. From an observed bimodality of the color distribution of globular star clusters surrounding

NGC 1316, Goudfrooij et al. (2004) concluded that a major merging event took place about 3 Gyr ago. In addition to this, smaller merging events have most likely occurred more recently (Schweizer, 1980; Horellou et al., 2001).

Located on the outskirts of the Fornax Galaxy Cluster, NGC 1316 is probably not a true member of the main cluster, centerd on NGC 1399, but belongs to a smaller sub-cluster (Drinkwater, Gregg & Colless, 2001). The distance to the galaxy has been estimated to 22.9 ± 0.5 Mpc by Goudfrooij et al. (2001) and 18.5 ± 1.3 Mpc by Jacoby, Phillips & Feldmeier (2004). For the sake of simplicity we shall here use the round value 20 Mpc, lying in between.

Optical observations early drew attention to the dark dust lanes in NGC 1316 and the similarity to the partly obscured galaxy NGC 5128 (Centaurus A) was pointed out by Evans (1949). Shklovskii & Cholopov (1952) and de Vaucouleurs (1953) associated NGC 1316 with the shortly before discovered, strong radio source, Fornax A. Later, Schweizer (1980) gave a thorough description of the general features of the distribution of dust in NGC 1316. Among other things, he pointed out that the dust patches seem to form two systems showing some degree of symmetry relative to the nucleus of the galaxy, one to the north-west and the other to the south-east. Within a distance of ~25″ (1″≈ 97 pc) from the nucleus, the predominant orientation of the dust patches is radial while at larger distances it is mainly azimuthal. With the entry of the Hubble Space Telescope (HST), the potential of obtaining sharper images of the dust pattern in NGC 1316 improved dramatically. Using the HST, *Wide Field Planetary Camera 2*, Grillmair et al. (1999) could image the central region of NGC 1316 in great detail.

In addition to stars and dust, NGC 1316 also contains ionized gas ([O II], [N II], [S II]) stretching in a broad band from ~35″ south-east of the nucleus of the galaxy to ~54″ north-west of the nucleus (Schweizer, 1980). The radial velocities of the gas, relative to the nucleus, range from about - 350 km $s^{-1}$ at the south-east end of the band to about 350 km $s^{-1}$ at the north-west end. Schweizer interpreted his observations in terms of a rotating, ionized gas disk seen rather much from the edge.

Later, CO observations performed by Sage & Galletta (1993) and Horellou et al. (2001) showed that molecular gas is present in NGC 1316 as well. The distribution of the CO gas roughly coincides with the distribution of the dark dust features. From the CO observations it could be settled that the velocity pattern of the molecular gas is fairly similar to that of the ionized gas but the span of the radial velocities is somewhat smaller. Horellou et al. obtained velocities, relative to the systemic, ranging from -280 km $s^{-1}$ at 32″ south-east of the nucleus to 180 km $s^{-1}$ at 72″ north-west of the nucleus. Under the assumption of a distance to the galaxy of 18.6 Mpc a total mass of the molecular hydrogen of $M_{H2} \approx 5\ 10^8\ M_{\odot}$, compatible with the CO measurements, was estimated. No atomic hydrogen could be detected ($M_{HI} < 1\ 10^8\ M_{\odot}$).

The aim of the present paper is twofold. A primary task is to study the dark features in NGC 1316 from a morphological point of view. For this purpose we will use a recent HST image of the central region of the galaxy showing the dark features with great definition. In particular, we will look for characteristic structures occurring repeatedly. A second task is to carefully consider some of the more complex and recurrent structures in an effort to shed light on their origin. In this matter, a model based on magnetized filaments is discussed.

## 2 Morphology of the dark features

A few years ago, the central region of NGC 1316 was imaged by the HST, *Advanced Camera for Surveys (ACS)* (see STScI-PRC2005-11). The image displays a wealth of dark and dusty features standing out in exquisite detail. A high-pass filter version of this image, emphasizing the small-scale features and suppressing the large-scale intensity variations in the galaxy, is shown in Figure 1. The figure reveals that a great number of dark filaments, blobs, and knots are present in the central region of NGC 1316. Some filaments are collected into long bundles or arcs while others are shorter and more complex. The dark features are essentially found within a circle of radius 75″, centerd at the nucleus of the galaxy. The nucleus, in turn, is placed at the center of the image in Figure 1. For the further study, comprising some of the more prominent structures, it is convenient to divide the features into six, more or less natural groups labelled from 1 to 6 (Figure 2).

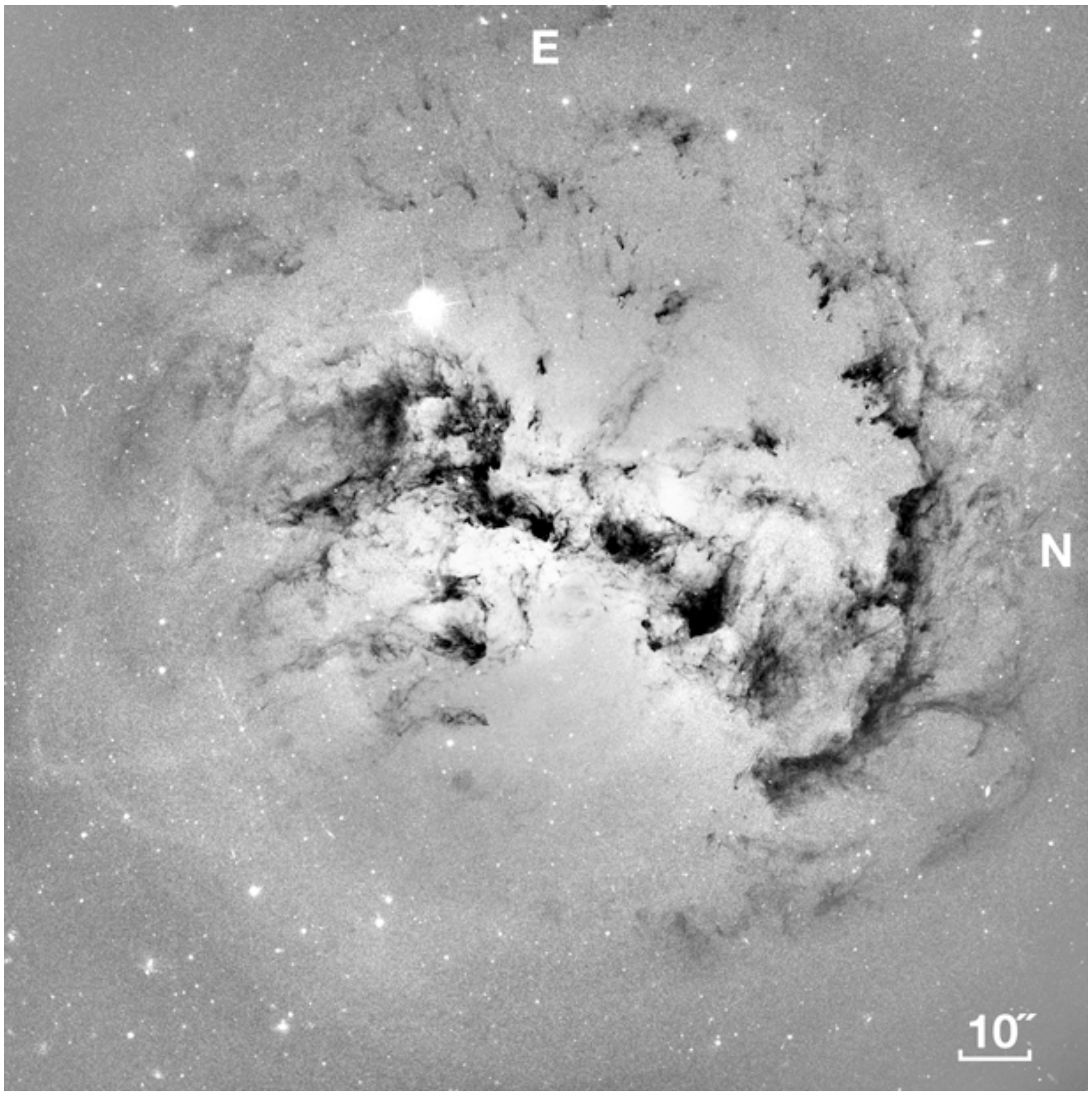

**Figure 1**. Central region of the giant, elliptical galaxy NGC 1316 containing a network of dark features. The original image (STScI-PRC2005-11), captured by the HST, *ACS*, has been processed by a high-pass filter suppressing the large-scale intensity variations of the galaxy and emphasizing the smaller, dark features. The nucleus of the galaxy is situated at the center of the image. Most of the dark features are markedly filamentary. A great number of dark, elongated structures protrude inwards, mainly from visible filaments. Image size, 2.′4 x 2.′4. East is up and north is to the right. Credit: The Hubble Heritage Team (STScI/AURA), NASA, ESA, P. Goudfrooij (2005). All the other images in the present paper, except for the image in Figure 3b, also benefit from STScI-PRC2005-11.

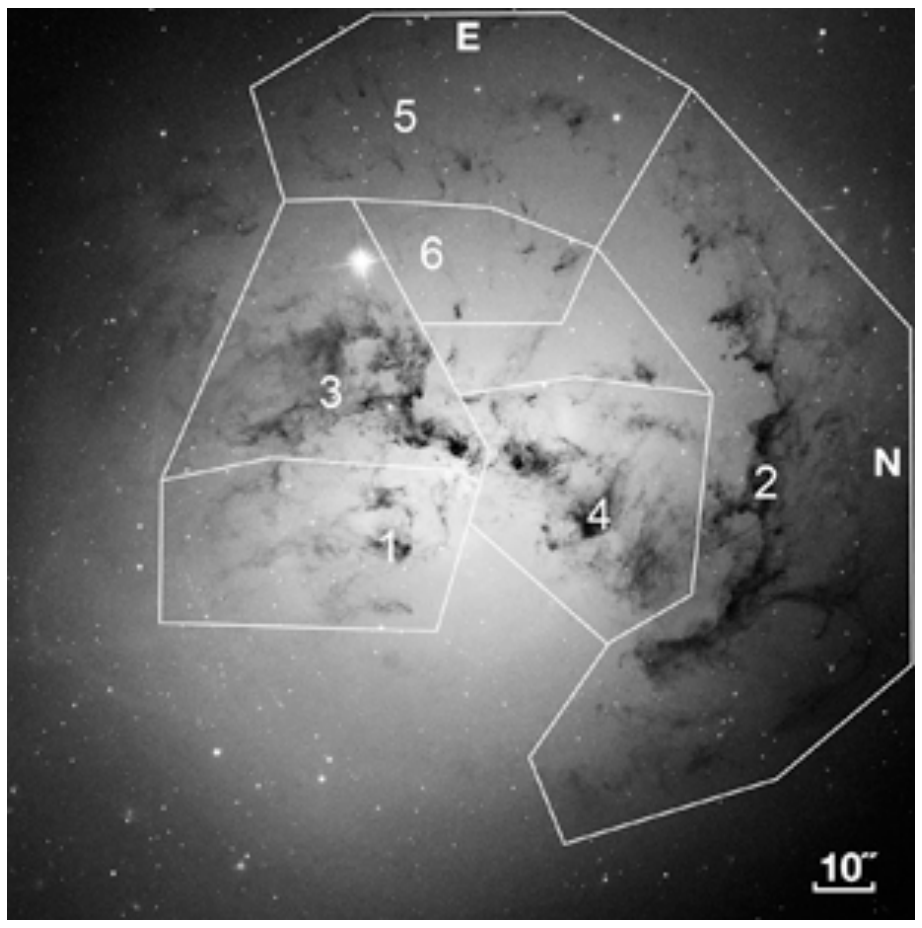

**Figure 2**. Map of identification of the various groups of mammoth trunks labelled from 1 to 6 and discussed in the text. The map is a greyscale version of the original HST, *ACS* image. The nucleus of the galaxy is at the center of the image and the orientation and angular size of the image are the same as in Figure 1.

*Group 1* includes three elongated cloud structures situated side by side (Figure 1). The clouds are stretching north by south and all of them are markedly filamentary. Their most massive parts (the heads) are in the north and they appear to connect to a web of thin filaments (widths, 0.″10 – 0.″25), mainly in the south. The easternmost and middle one of the clouds probably consists of more than one elongated unit. The westernmost cloud, shown in some more detail in Figure 3a, is made up of thin filaments appearing to be twined around each other. Interestingly, the cloud is strikingly similar to one of the elephant trunks in the Rosette Nebula (Figure 2 in Carlqvist, Gahm & Kristen, 2002, hereafter CGK02), or to be exact, its mirror image, reproduced in Figure 3b. Without knowing anything about the origin of the object shown in Figure 3a, *it would be hard to assert, only by visual inspection of the image, that the object does not represent an ordinary elephant trunk in an H II region in the Milky Way Galaxy*. Still, the two objects in Figure 3 are of very different dimensions. The projected length of the filamentary cloud in NGC 1316 is ~1.0 kpc while the projected length of the Rosette trunk is about 0.9 pc.

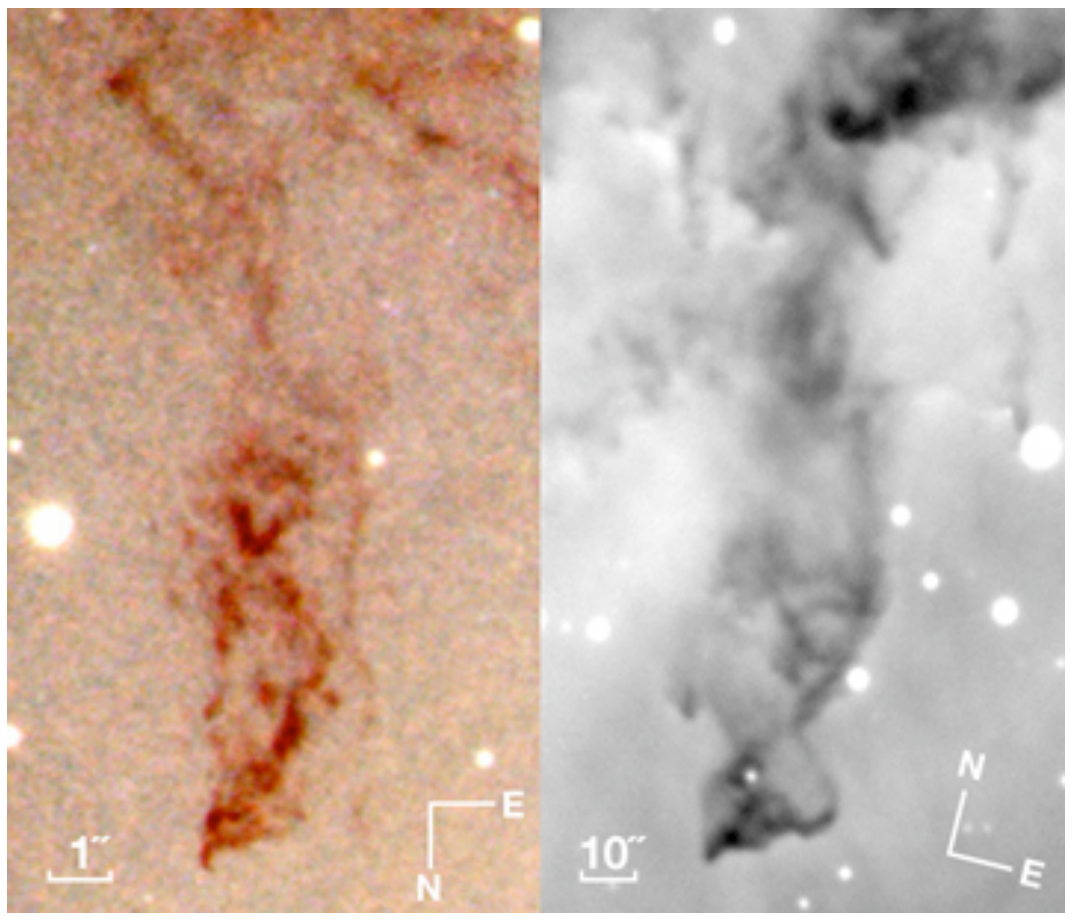

**Figure 3**. **a**) (Left panel): Elongated structure belonging to Group 1 and situated ~25″ west-south-west of the nucleus of the galaxy. The structure is made up of filaments, appearing to be twisted about each other. In the north, the filaments form a closed structure. In the south, they connect to a network of thin threads. The projected length of the structure is about 1.0 kpc while the widths of the individual filaments amount to ~10 – 25 pc. Image size, 8″ x 14″. South is up and east is to the right. **b**) (Right panel): Mirror image of an elephant trunk in the northern part of the Rosette Nebula. The trunk consists of winding filaments. At an adopted distance of 1.6 kpc, the projected length of the trunk is ~0.9 pc. The image was captured in the H$\alpha$ line by means of the 2.6 m Nordic Optical Telescope (CGK02). It is centerd at $\alpha = 6^h\ 31^m\ 36.^s5$ , $\delta = 5°\ 27.'1$ (Epoch 2000.0) and covers 1.′5 x 2.′6. The similarity of this trunk and the structure shown in panel a) is obvious.

To the north of the middle and easternmost of the three elongated clouds are a few, fairly weak, filamentary structures pointing eastwards towards the nucleus of the galaxy. The structures have their heads in the east and seem to be connected to the web of thin filaments in the south, just as the three elongated clouds.

*Group 2* consists of a filamentary system containing a prominent arc. The arc is azimuthally oriented and runs through half of the north-eastern quadrant and half of the north-western quadrant at a distance of about 50″ from the nucleus of the galaxy. Consisting of more than two filaments, the arc shows signs of being twisted. A number of dark, elongated structures are protruding ~2″– 8″ from the filaments in the arc and point inwards, towards the central parts of the galaxy. Typically, the deviations from the radial direction are in the interval 0°– 50°. Many of the structures resemble elephant trunks in galactic H II regions. Grillmair et al. (1999) observed that some of the structures in the north-eastern quadrant are similar to the well-known columns in the Eagle Nebula (M 16) (Hester et al., 1996).

Most of the structures jutting out from the arc are shaped like the letter Y with the stem of the Y roughly pointing towards the central parts of the galaxy (Figure 4). The outer parts of the structures (the legs) are often split and connect to filaments in the arc. In some of the protruding structures there are signs of two or more filaments winding around each other. In most of these structures there is a mass condensation near the end facing the nucleus. About 45″ to the north of the nucleus, a conspicuous Y-shaped structure protrudes southwards from the arc (Figure 5). The structure has two legs in the north and appears to be twisted. A feature that immediately attracts the attention is the split head. Heads of such a shape are not uncommon among elephant trunks (e.g. the Wrench Trunk in the Rosette Nebula and the Stag-Beetle Trunk in IC 1805 (Carlqvist, Gahm & Kristen, 2003, hereafter CGK03)). The projected length of the whole structure is ~700 pc.

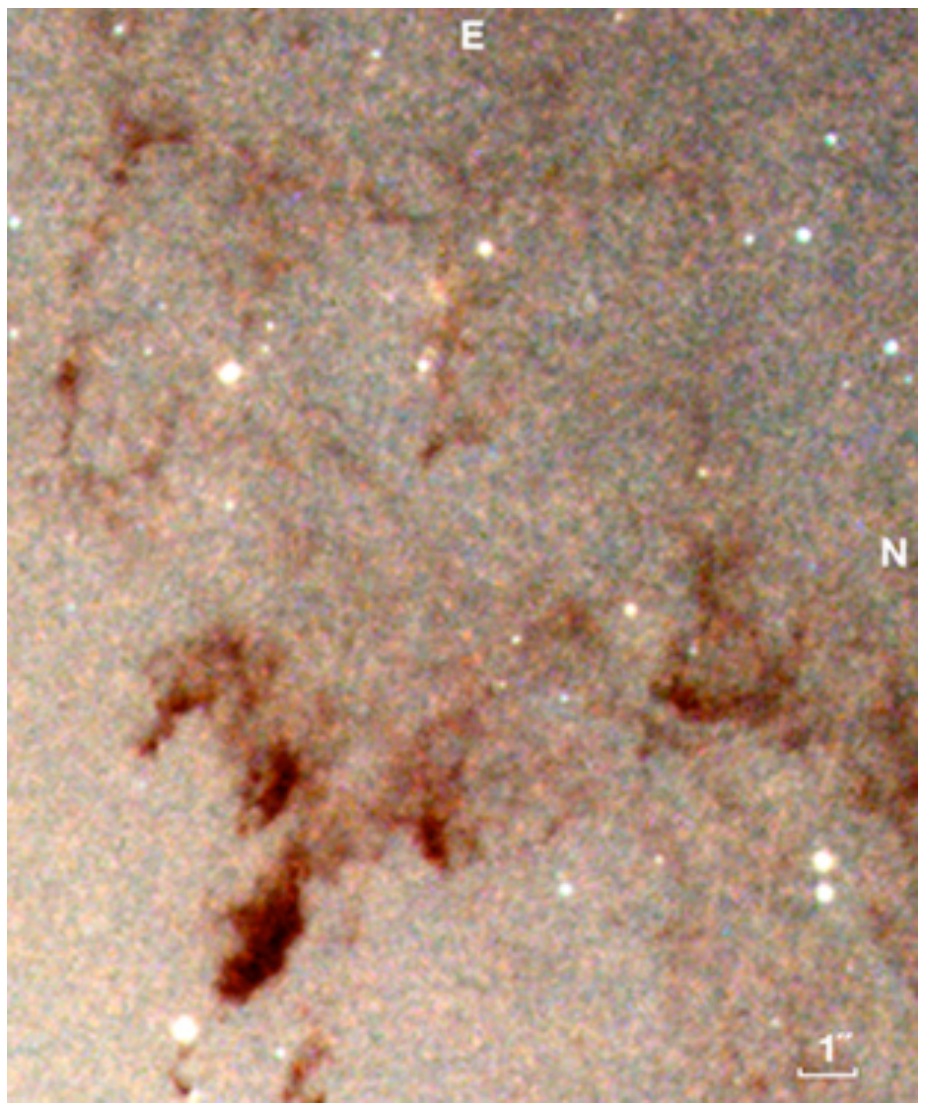

**Figure 4**. Dark, protruding structures belonging to Group 2 and situated about 50″ north-east of the nucleus. The massive structures seen on this image are not pointing exactly towards the nucleus but ~20° to the north of it. Most of the structures have legs connecting to the filaments and some are wiggly indicating that they are made up of twined filaments. The wiggly structures are, however, only marginally resolved. To be noted also, are the thin tendrils which hang down west-south-west-wards from transverse, slender filaments in the upper, left-hand part of the image. Image size, 15″x 18″. East is up and north is to the right.

Some distance further to the west, a bundle of filaments separates from the arc and runs to the north (Figure 1). The arc and the bundle together form a structure that may be described as Y-shaped. This structure is, however, considerably larger than the Y-shaped structures mentioned above. A close inspection of the bundle indicates that the filaments are intertwined. Somewhat further out, the bundle turns to the west and forms an azimuthally oriented, more diffuse band. The band seems to run all the way to the border of the south-western quadrant while dispersing. East of the bundle, and to the north of the prominent arc, a number of weak and fuzzy filaments are stretching north-east by south-west. Only a few of these show a clear V- or Y-shape.

*Group 3* is represented by a system of wavy filaments going southwards from a pronounced mass condensation projecting next to the nucleus of the galaxy (Figure 1). Further out, it bends westwards while reaching out to distances of more than 40″ from the nucleus. The system is composed of a great number of thin filaments, out of which some show a sinusoidal shape. The widths of the filaments are mainly in the interval 0.″10 – 0.″20. In the middle of the system, about 12″ from the nucleus, there is a thick structure putting out towards the east. At the first glance, it might seem that we are also here dealing with a Y-shaped structure, but this is probably an illusion. A detailed inspection of the HST, *ACS* image suggests that the protrusive structure constitutes an independent unit having a shape similar to the letter V.

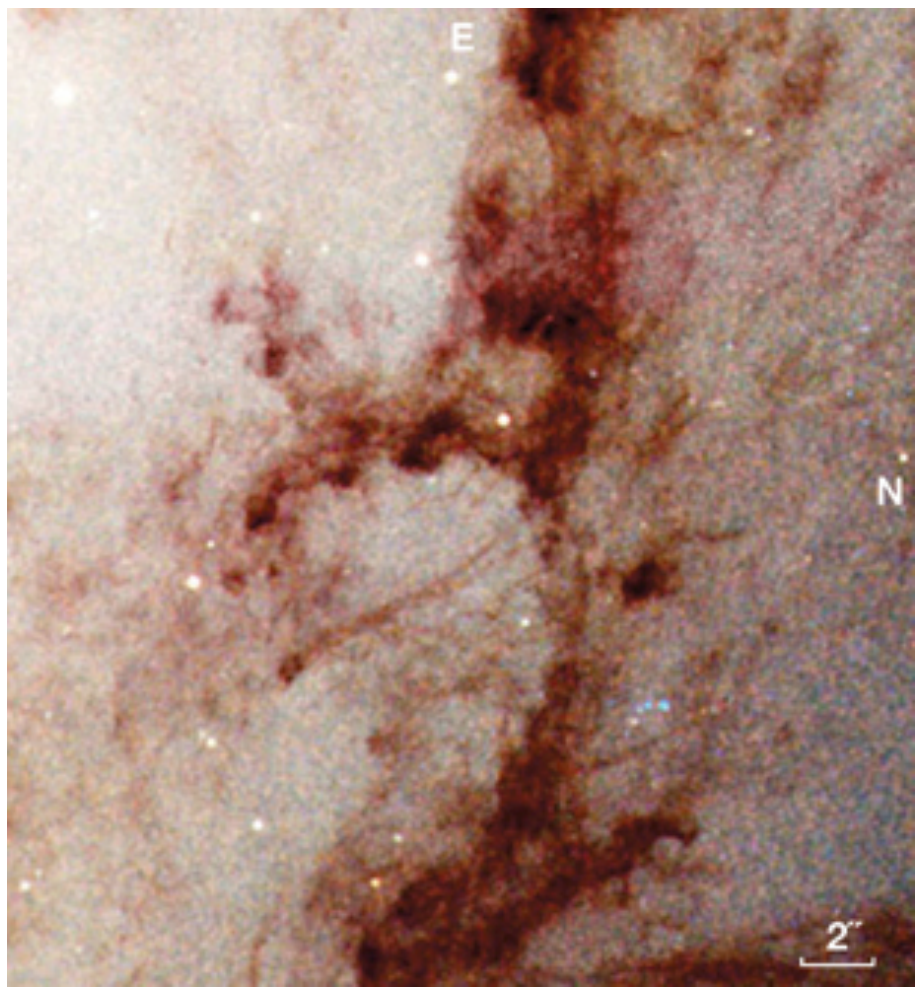

**Figure 5**. A protruding, Y-shaped structure belonging to Group 2, situated about 45″ north of the nucleus. The structure has a split head to the left and shows clear signs of being twisted. To the right it is connected to the prominent arc by two main filaments forming the legs. A much thinner, two-legged structure, resembling a pine-needle, is to be seen in the lower half of the image. The mass condensation is to the left and the two legs, forming a much drawn out V, goes to the right. Image size, 24″x 26″. East is up and north is to the right.

*Group 4* is composed of three dark patches being lined up to the north-north-west of the nucleus and reaching out to a distance of about 40″ from the nucleus. Each of the patches consists of a network of thin filaments, which may give a rather chaotic impression. A closer study, however, suggests that the innermost patch is a superposition of two comet-like, filamentary structures with the tails oriented towards east-south-east and north-north-east. The dark patch in the middle, resembling a ray, has both head and tail. The tail is wavy and oriented towards east-north-east. The outermost patch, finally, has no obvious head or tail but is filamentary with the filaments mainly stretching east-north-east by west-south-west. Immediately to the east of the patches, but also in between them, there are several small, filamentary structures presenting V- and Y-shapes.

*Group 5* consists of a number of relatively small, comet-like objects situated about 50″–70″ east of the nucleus (Figure 6). The objects are not randomly spread over the area but stand out in fairly regular rows against the background. Some of the objects are distributed along an azimuthally oriented, mainly invisible arc at distances of ~50″– 55″ from the nucleus. About 10″ outside this imaginary arc there is another row of similar objects. The rows seem to be connected through a number of intertwined filaments east-south-east of the nucleus. The inner, imaginary arc appears to be a continuation of the prominent arc in Group 2. Together, they form a gigantic, azimuthally oriented structure covering more than 160°. The comet-like objects all point inwards in the galaxy. It is to be noticed that they do not point exactly to the nucleus but have a systematic northward tilt of typically 10°– 40°. The projected lengths of the larger structures in Figure 6 are ~ 200 – 500 pc while their widths are ~ 40 – 90 pc. Most of the structures are Y-shaped or V-shaped and some appear to be twisted next to their heads. Also in this group, many of the structures show a considerable resemblance to elephant trunks.

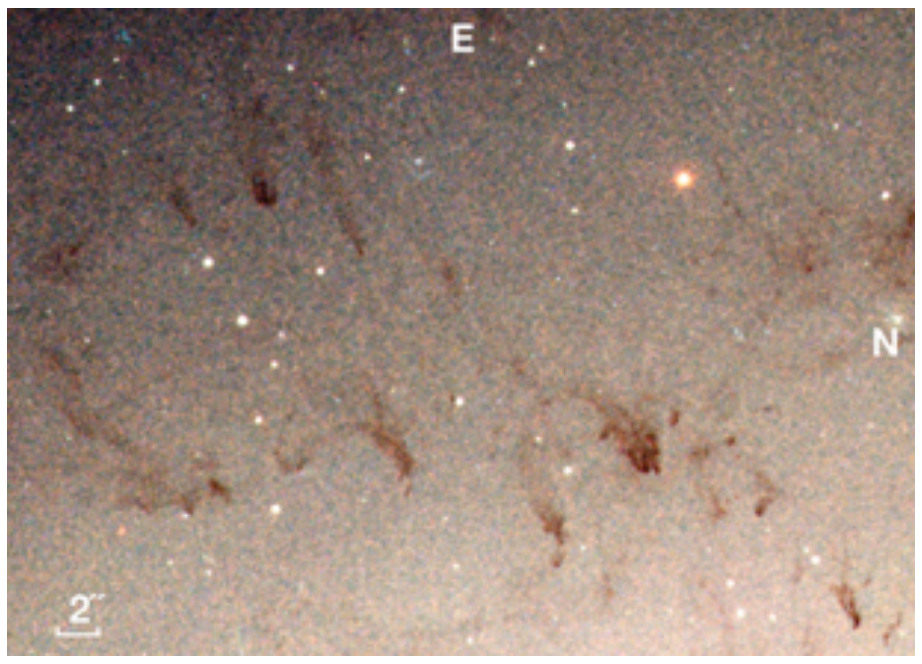

**Figure 6**. Rows of relatively small (~2″– 5″), V- and Y-shaped structures in Group 5 situated ~50″– 70″ east of the nucleus of the galaxy. The structures are oriented inwards and most of them point ~10°– 40° to the north of the nucleus. Some structures show signs of being twisted. Image size, 40″x 29″. East is up and north is to the right.

*Group 6* finally, includes several faint and apparently insignificant, comet-like structures situated about 25″– 45″ east of the nucleus (Figure 1). Parts of the tails of some of the structures can be seen just above the lower edge of Figure 6. Although faint, they are still of great interest since some of them appear to be Y-shaped, a few with a twined stem. A possible reason for their faintness may be that they are located deeper in the galaxy than most of the other dark structures.

The study of the dark features in the central region of NGC 1316 shows that it is possible to sort out a few typical and recurrent kinds of dark structure. One prominent property of the dark features is that most of them are filamentary. Both very long filaments and shorter filamentary parts are present. A number of the filamentary parts are probably connected with each other although the amount of dust is locally too low to reveal this. Many of the filaments are azimuthally oriented, particularly in the outer part of the dusty region, but other orientations exist as well.

Another kind of recurrent feature is constituted by the relatively short structures protruding inwards from the filaments. Many of these structures are strikingly similar to elephant trunks in H II regions, although much larger. Most of the structures have lengths in the interval 1″– 10″ and often they are V- or Y-shaped, just as many elephant trunks. To a large extent, thin filaments with widths of about 0.″10 – 0.″25 build up the structures. In some of the Y-shaped structures, there are indications of that the stem of the Y consists of two main filaments being wound around each other. In most cases, however, the resolution is insufficient to clearly reveal the internal structure of the stems. Since, from a morphological point of view, many of the dark, protruding structures in NGC 1316 are so similar to elephant trunks, but much larger, we will term them *mammoth trunks*.

The study performed, clearly shows that most of the mammoth trunks roughly point inwards, towards the central parts of the galaxy (Figures 1 and 7). The spread of their orientations with respect to the radial direction is, however, considerable and a few trunks are even nearly perpendicular to radius vector. Thus, there is an appreciable scattering of the deviations of the trunks from the direction pointing towards the nucleus of the galaxy.

If instead, we abandon the radial direction as a norm, the orientations of the mammoth trunks show a more regular pattern, although in more limited regions. A closer look at the orientations of the mammoth trunks in Figure 7 reveals that there exist fairly large sub-regions, within which the orientations are well-ordered and in some cases nearly parallel with each other. One such sub-region is constituted by most of the trunks in Groups 5 and 6 where the trunks are oriented towards west-north-west. Likewise, the greater part of the trunks in Group 2, but for the small aggregation of trunks shown in Figure 4, form another sub-region with the trunks pointing towards south-south-west. The eastern end of the major arc may also be a member of this sub-region. A further example is given by the three main trunks in Group 1, oriented towards the north, perhaps supplemented by the wavy filamentary system in Group 3. Besides the sub-regions with well-ordered trunk orientations, there are also a few spots where the orientations of the trunks seem to cross each other. However, the well-ordered sub-regions dominate.

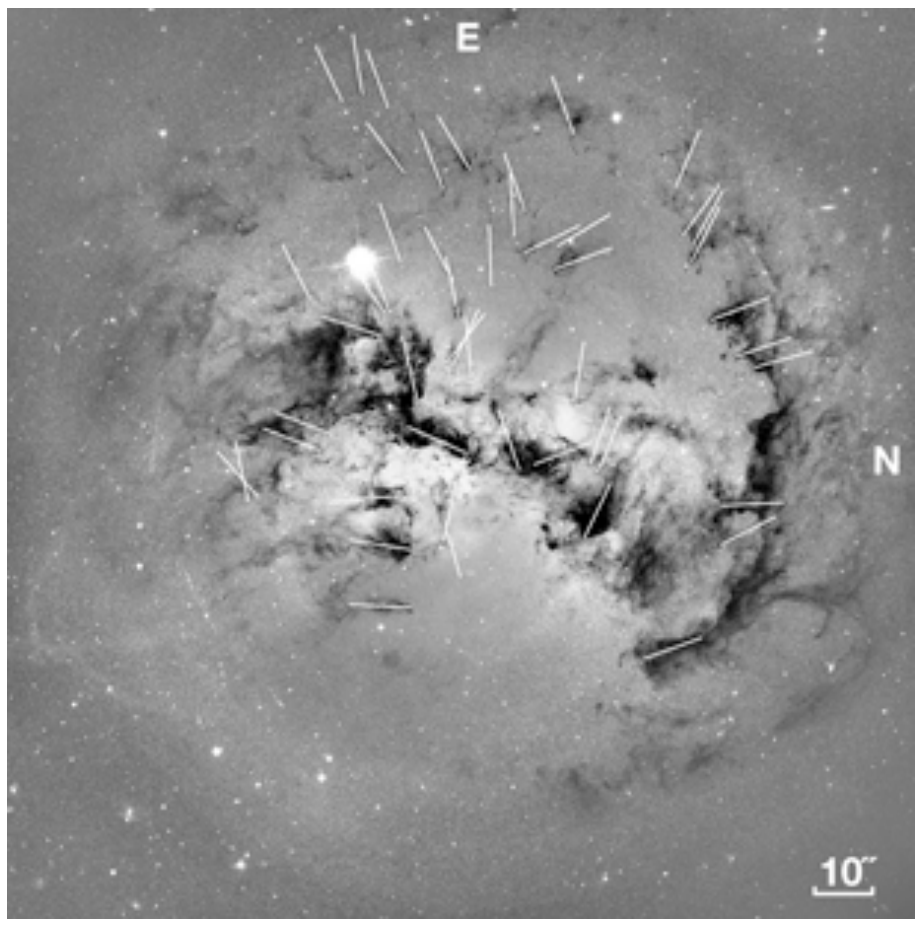

**Figure 7**. Orientations of some of the more distinct mammoth trunks in NGC 1316, indicated by short lines. The ends of the lines, closest to the nucleus of the galaxy, mark the heads of the trunks. The orientations are found to form fairly regular and well-ordered patterns within several sub-regions. The nucleus of the galaxy is at the center of the image and the angular size and orientation of the image are the same as in Figure 1.

From the similarities of the mammoth and elephant trunks, one might suspect that the basic mechanism behind both of these kinds of object should also be similar. In the search for the origin of the mammoth trunks, we will therefore have a closer look at models proposed for elephant trunks.

# 3 Properties and origin of elephant trunks

Elephant trunks are dark, elongated structures seen in silhouette against bright H II regions (e.g. Figure 4). The H II regions are surrounded by molecular clouds and contain in their central parts luminous OB stars responsible for the heating and ionization of the regions. In general, the elephant trunks point inwards from the rim of the regions towards the OB stars. Some of the trunks consist of a fairly loose network of filaments and, hence, give the impression of being veil-like. In such trunks, the individual filaments are relatively easy to trace and often seem to form spirals along the trunk (e.g. the Wrench Trunk, (CGK03)). Other trunks again, represent more compact pillars where the transparency is low or non-existent. Even if it is not possible to trace individual filaments in such trunks, there may still be indications of the presence of filaments. In particular, filaments may be discerned near the edges of the pillars. In some cases there also exist diagonal structures on the face of the trunks indicating helical geometry.

Since long, elephant trunks have been considered to be caused by the Rayleigh-Taylor instability as hot, expanding plasma inside the H II region sweeps up the denser gas in the surrounding molecular cloud into a shell (Spitzer, 1954; Frieman, 1954). However, with the discovery of filamentary and twisted elephant trunks (CGK02) it has turned out hard to see how this theory can properly describe the trunks. In order to account for the properties discovered, a new theory, based on the double helix mechanism, has been proposed (CGK02; CGK03).

In the new theory, magnetized filaments, sometimes termed magnetic ropes (Babcock, 1961), constitute a basic element. The filaments are parts of the molecular cloud surrounding the H II region and consist, just like the cloud, of cold, low-ionized, and dusty gas. Although the degree of ionization is low ($\sim 10^{-7}$– $10^{-4}$, Hjalmarsson & Friberg, 1988; Black & van Dishoeck, 1991), the gas still constitutes a plasma. Furthermore, the magnetic Reynolds number $R_m$ is much larger than unity implying that the magnetic field lines are well frozen-in to the gas for long time periods (e.g. Alfvén & Fälthammar, 1963; Spitzer, 1978; Ruzmaikin, Shukurov & Sokoloff, 1988). As the hot plasma inside the H II region expands, surrounding molecular gas is swept up in a shell moving outwards (snow-plough effect). The filaments are, of course, also influenced by this process. If a filament contains matter of low enough density, it will simply be swept up by the shell and move along with it as an inherent part. But, if the filament contains a sufficiently large mass condensation, the latter will lag behind the rest of the filament because of its larger inertia. In case the magnetic field lines inside the filament are only moderately twisted, or not twisted at all, the filament is stretched and attains the shape of a V with the mass condensation situated at its tip. The tip of the V faces the central, luminous stars while the upper parts of the V connect to the rest of the filament in the expanding shell. If the magnetic field lines inside the filament are twisted beyond a certain critical limit, e.g.

as a result of convective motions in the medium surrounding the filament (cf. Babcock, 1961), the very filament next to the tip is twisted into a double helix so that the filamentary structure resembles a Y. The stem of the Y is identified with an elephant trunk.

To find out whether a given twist of the magnetized filament will lead to a V-shape or Y-shape of the filament, one can benefit from two principles. The first principle implies that the filamentary system should tend towards an energy minimum while the second principle prescribes that the helicity of the magnetic field, $H = \langle \boldsymbol{A}\cdot\boldsymbol{B} \rangle$, has to be constant (Woltjer 1958; Moffat 1969). $\boldsymbol{A}$ is here the vector potential for the magnetic field $\boldsymbol{B}$, and $\langle\ \rangle$ stands for a volume average over the flux tube. The critical limit for transition from V-shape to Y-shape is found to be reached when the azimuthal magnetic field component in the outer layers of the filament becomes of the same order as the axial component. A more thorough account of the double helix mechanism is given in CGK03.

The double helix mechanism can also be illustrated by means of a simple analogy model (CGK02). The model consists of a bundle of elastic strings, firmly attached to the ends of a rod, and a weight hanging down from the middle of the bundle. Here, the bundle of strings represents the magnetic field lines frozen-in to the filament (Faraday, 1839–55; Alfvén, 1950) while the gravitational force acting on the weight simulates the inertia force of the mass condensation in the filament. The attachment of the bundle of strings to the rod imitates the connection of the filament to the expanding shell. If now, the bundle of strings is only slightly twisted, or not twisted at all, the bundle hangs down in the form of a V with the weight at its tip. If instead, the bundle is twisted beyond a certain critical limit, the bundle forms a double helix near the tip of the former V so that the overall shape resembles a Y (e.g. Fig. 3 in CGK02). The twisting of the bundle into a double helix is suggestive of what happens to the rubber string in a model plane when the string is too much twisted.

Since the magnetic field is well frozen-in to the gas, the formation of a double helix in an elephant trunk must necessarily involve a rotation of the trunk about its long axis during some period of its existence. Recently, CO observations have revealed that such a rotation indeed occurs in elephant trunks (Gahm et al., 2006). Rotation was found in four out of four investigated trunks.

## 4 Model of the mammoth trunks

### *4.1 Approach to a model*

The similarity of the dark mammoth trunks in NGC 1316 and elephant trunks in H II regions strongly suggests that the mechanisms behind both of these kinds of object should be similar. When searching for a model of the mammoth trunks, it is therefore important to recall that the mammoth trunks, just as the elephant trunks, present both V-shaped and Y-shaped, filamentary structures and in some cases even twined structures. From these observations, one may conclude that the same drawbacks that are hampering the Rayleigh-Taylor model of elephant trunks should also affect a similar kind of model applied to the mammoth trunks. By contrast with this, the double helix mechanism is well adapted to account for V- and Y-shapes, including twined structures. Hence, it is natural to consider a model of the mammoth trunks, based on this mechanism.

*4.2 The filaments*

A basic element in the double helix mechanism is the magnetized filament forming a more or less twisted rope. There are plenty of such filaments in the central region of NGC 1316. Many of these appear to be relatively short while others are longer like the prominent arc in Group 2 ranging over 80″, which corresponds to at least 8 kpc. As indicated by the CO observations performed by Sage & Galletta (1993) and Horellou et al. (2001), the dark filaments mainly consist of a dusty molecular gas, predominantly $H_2$. Most likely, the filaments constitute part of the remnants of a small and gas-rich, spiral galaxy, which collided and merged relatively recently with the progenitor of NGC 1316.

It must be considered remarkable that so many of the filamentary structures have been preserved as coherent units in NGC 1316 in spite of their violent history. The only reasonable explanation of this robustness seems to be that the filaments are to a great extent controlled by strong magnetic fields. This is not surprising, since it is well known that strong magnetic fields exist in the arms of spiral galaxies. For instance, the magnetic fields in molecular clouds in the arms of the Milky Way Galaxy typically are of the order $B \approx 1 - 10$ nT = $10 - 10^2$ μG for cloud densities of $10^8 - 10^{10}$ m$^{-3}$ = $10^2 - 10^4$ cm$^{-3}$ – the denser the cloud the stronger the field (e.g. Troland & Heiles, 1986; Myers & Goodman, 1988; Valleé, 2003). This implies that the energy density of the magnetic field is comparable with the kinetic energy density of the gas, including turbulent motions. Owing to the fact that the magnetic field is divergence-free, div $\boldsymbol{B} = 0$, the filaments in NGC 1316 tend to keep together as coherent units, although probably deformed and locally compressed or diluted. Just as is the case for the elephant-trunk filaments, the magnetic field lines must be well frozen-in to the dark filamentary medium in NGC 1316.

Before considering a model of the mammoth trunks more in detail, it may be appropriate to say a few words about the interpretation of the dark, elongated structures in NGC 1316. When describing the structures in Section 2, we have mostly used the terms *filament* and *filamentary*. In older literature, one may see the dark features in NGC 1316 described as *patchy* and sometimes even as *chaotic*. This was quite natural at that time but with the entrance of the detailed HST images it has become clear that most of the dark features keep together in long and generally tangled threads and bands. A question that naturally presents itself is, could the thin, elongated features possibly be sheets, e.g. in the form of shock fronts or shells? We do not think so. There are several arguments in support of this view. Since the thin features are mostly discrete and since, furthermore, there is little variation of their widths (e.g. see Figure 3), the sheets would have to be seen nearly edge-on, an extremely unlikely situation. Furthermore, many of the recurrent and rather complicated, filamentary structures are not compatible with shock fronts or shells. Hence, the dark elongated structures are interpreted as filaments.

*4.3 The external force*

Since the mammoth trunks in NGC 1316 are not randomly arranged but, on the whole, fairly consistently oriented inwards, there must exist some external, adjusting force acting on the filaments. This force can, however, not be of the same nature as the external force acting on elephant trunks. There is simply no evident, expanding shell in the region in question to which filaments can be tied. Instead, the origin of the external force is likely to be sought in the relative motion of the filaments and an ambient gaseous or plasma medium. Two fundamentally different options are here at hand. Either a fast wind blows outwards, past the filaments, from the central part of the galaxy, or do the filaments move inwards through a fairly stationary medium.

As regards the first option, a sufficiently strong galactic wind with its source near the center of the galaxy could very well have the capacity to influence the filaments so that parts of them would be aligned with the wind. However, it is hard to see how an almost radially blowing wind could give rise to the orientations found among the mammoth trunks (Figure 7). As pointed out in Section 2, there is a considerable spread of the orientations of the trunks with respect to the radial direction. A few trunks are even nearly perpendicular to the radial direction. Of course, one might imagine a situation where several wind systems emanate from different sources in the central region of the galaxy. Disregarding the difficulty of finding a reasonable physical explanation of such a complex system, it is doubtful whether it can offer a satisfactory explanation. A main reason for this is that the orientations of the mammoth trunks are fairly well-ordered within separate sub-regions and do not diverge as one would expect if the trunks were acted upon by local winds.

The second option implies that infalling magnetized filaments are acted upon by a dynamic pressure as they make their way inwards through the target galaxy. It is well known that the collision and merging of two galaxies can lead to rapid inflows of molecular gas towards the central parts of the resultant galaxy due to gravitational torques (Tomre & Tomre, 1972; Negroponte & White, 1983; Sanders, Scoville & Soifer, 1991; Barnes & Hernquist, 1991, 1996, and references therein). In consequence of this, dark filamentary matter is expected to fall down through the potential well of NGC 1316. The filamentary matter is not thought to move as a rigid body but rather as filamentary pieces, coupled together by the magnetic field. Such pieces may correspond to the sub-regions found, containing mammoth trunks with fairly well-ordered orientations. Some of the filamentary pieces are likely to exist in more than one layer and such an arrangement might give rise to limited regions containing trunks with crossing orientations.

From the arguments given above, it is clear that the second option seems to be better suited than the first one in accounting for the trunk orientations observed in NGC 1316. One may of course wonder why an elliptical galaxy like NGC 1316 should house an ambient plasma medium, like the one suggested, in its central region. A possible answer to this question is discussed in Section 5.3. For the time being, we shall be content with pointing out that the CO observations performed by Sage & Galletta (1993) and Horellou et al. (2001) may offer some hints on the speeds involved in the motion of the filaments relative to the galaxy. Typical radial velocities of the molecular gas, relative to the system, are of the order of one to a few hundred km $s^{-1}$ (Section 1). Taking the two unobserved velocity components into account, the velocity of the molecular gas, relative to the galaxy, must be still higher and thus even more effective in contributing to the formation of the mammoth trunks.

*4.4 Formation of mammoth trunks*

From the above discussion a model of the mammoth trunks emerges. As an inspection of the HST, *ACS* image of NGC 1316 reveals, the dusty filaments do not possess a constant greyness but consist of weaker parts interspersed with darker parts. Hence, it is clear that the filaments are not uniform but have a mass density that varies along their axes. Depending on the magnitude of the mass density, different parts of a filament are expected to react differently upon the external force. If the inhomogeneity along the filament involves a sufficiently pronounced mass condensation, the filament will be pulled apart sideways, since the mass condensation is less braked than the more dilute parts of the filaments. Still, the filament must form a coherent unity because of the inherent magnetic field.

To illuminate this process, we consider a dark, magnetized filament of radius $r_f$ containing a gas that mainly consists of $H_2$ molecules. Inside the filament, the magnetic field lines are supposed to be of helical geometry with a moderate twist. The filament falls through an ambient plasma consisting of hydrogen atoms, protons, and electrons. As a result of the relative motion $\boldsymbol{v}_r$, the filament is subject to a dynamical pressure (ram pressure). In the simple situation where the filament is perpendicular to $\boldsymbol{v}_r$, the dynamical pressure gives rise to a force per unit length of the filament of

$$F_{dyn} = 2 r_f n_p m_p v_r^2 \quad . \tag{1}$$

Here, $n_p$ is the density of the ambient plasma (the sum of the densities of the protons and hydrogen atoms) while $m_p$ is the proton mass. The force contributes to the acceleration of the filament by the amount

$$a_{dyn} = \frac{F_{dyn}}{M_f} \quad , \tag{2}$$

where

$$M_f = \pi m_g n_g r_f^2 \tag{3}$$

is the mass per unit length of the filament. The quantities $m_g$ and $n_g$ represent the mass and density of the molecules in the filament, respectively. In most parts of the filaments, the density is supposed to be of the order $n_{g0} \approx 10^2$ $cm^{-3} = 10^8$ $m^{-3}$. If now, there is a local mass condensation in the filament with a density of, say, $n_{g1} \approx 10^3$ $cm^{-3} = 10^9$ $m^{-3}$, this matter will be subject to an acceleration (braking) that is smaller than the acceleration (braking) of the surrounding filament. The difference of the two accelerations is

$$\Delta a_{dyn} = \frac{F_{dyn}}{\pi m_g r_f^2}\left(\frac{1}{n_{g0}} - \frac{1}{n_{g1}}\right) \quad . \tag{4}$$

Since $n_{g1} >> n_{g0}$, we have

$$\Delta a_{dyn} \approx \frac{F_{dyn}}{\pi m_g n_{g0} r_f^2} \quad . \tag{5}$$

During the time period $\tau$, when the filament is exposed to the dynamical pressure, the difference in acceleration will cause the filament to be locally drawn out laterally. As part of the filament is extended, forces due to magnetic tension and gravity also come into play. In general, however, these forces will only cause the filament to be bent less sharply than it would be in the absence of the forces. Only if the mass condensation is very small (of the order of $r_f$ or smaller) the extension is considerably reduced. In more normal situations comprising a larger mass condensation, the result is a V-shaped extension with gently rounded connections to the rest of the filament and a length of

$$\Delta \ell \approx \Delta a_{dyn} \frac{\tau^2}{2} \quad . \tag{6}$$

By means of equations (1) and (5), equation (6) can be transformed to

$$\Delta \ell \approx \frac{n_p}{n_{g0}} \frac{m_p}{m_g} \frac{(\tau v_r)^2}{\pi r_f} \quad , \tag{7}$$

which in turn, with $m_g \approx 2m_p$ , yields

$$\frac{n_p}{n_{g0}} \approx \frac{2\pi r_f \Delta \ell}{(\tau v_r)^2} \quad . \tag{8}$$

Since $\tau$ represents the time of exposure to the dynamical force, the quantity $\tau v_r$ should be of the order of the linear dimension of the region containing the dark features. Hence, we choose $\tau v_r$ = 150″ ≈ 15 kpc. With typical relative speeds of $v_r$ = 100 – 300 km s$^{-1}$, the exposure time turns out to be $\tau \approx 5\ 10^7 - 1.5\ 10^8$ years. The density of the ambient plasma can now be estimated using equation (8). In Section 2, the lengths of the mammoth trunks were found mainly in the interval 1″ – 10″. Taking a mammoth trunk of intermediate length $\Delta \ell$ = 5″ ≈ 500 pc and putting $r_f$ = 0.″1 ≈ 10 pc, we obtain the ratio of the densities $n_p/n_{g0} \approx 1\ 10^{-4}$. With the particle density considered above, $n_{g0} \approx 10^2$ cm$^{-3}$ = $10^8$ m$^{-3}$, the density of the ambient plasma required to stretch the trunks appropriately is $n_p \approx 1\ 10^{-2}$ cm$^{-3}$ = $1\ 10^4$ m$^{-3}$.

If the ambient plasma is assumed to completely fill a sphere containing the dark features, i.e. having a radius of ~ 7.5 kpc, the total mass of the plasma turns out to be ~ 3 $10^8$ $M_{\odot}$. This mass is of the same order as the total mass of $H_2$ in NGC 1316, estimated from CO observations (Section 1). It is likely, however, that the ambient plasma is not homogeneously distributed in the inner region of the galaxy but possesses some degree of clumpyness. With a filling factor of less than 100%, the mass of the ambient plasma is likely to be reduced.

In the process described above, resulting in a V-shaped trunk, it was assumed that the magnetic field lines in the filament were only moderately twisted. For larger twists, exceeding the critical limit for double helix formation, the result is instead a Y-shaped trunk where the stem of the Y consists of two main filaments winding around each other. The mass condensation is situated at the bottom of the stem, while the upper parts of the Y, being more diluted, connect to the rest of the filament. In the model considered, the V- and Y-shaped mammoth trunks are expected to point with their heads in essentially the same direction as the dark structures move, relative to the ambient medium. If the ambient medium is relatively fixed and the dark filaments mainly move inwards in the galaxy as presumed, the trunks should also have a general tendency of pointing inwards.

## 5 Discussion

### *5.1 A different aspect of the double helix formation*

It is of some interest to note that the Y-shaped structure, prescribed by the double helix model, is constituted by a filament that is twisted on two different levels. First, the very filament is twisted so that the magnetic field lines in it form spirals. Secondly, the two parts of the filament forming the stem of the Y are twisted around each other, although in the opposite sense as compared with the twisting within the filament. The formation of the double helix in the stem may be considered a good example of self-organization taking place when the filamentary system has been forced far away from equilibrium. In this view, the transition from V-shape to Y-shape is signified by a point of bifurcation (equivalent to the critical limit of the twisting) separating two levels of fundamentally different organization.

### *5.2 Radio jet – dark filament interaction?*

When discussing the physical reason for the external force in Section 4.3, we favored a mechanism comprising an infall of the dark filaments through a more or less stationary plasma to an omnidirectional plasma wind blowing outwards from the central part of the galaxy past the filaments. Now, there actually exists an outflow from the central part of NGC 1316, although an outflow of a very special kind. The outflow is constituted by two, fairly well collimated jets belonging to the radio galaxy Fornax A. The jets emanate from the radio core of Fornax A, coinciding with the optical nucleus of NGC 1316 to within 0.″5 (Geldzahler & Fomalont, 1984). With observed lengths of ~ 30″– 40″, the jets are situated well inside the region containing the dusty filaments in NGC 1316. The jets do not appear to be connected with the two main radio lobes, which reach out to ~ 25′ from the radio core (Ekers et al., 1983). Radio maps obtained by Geldzahler & Fomalont (1.5 GHz, 4.9 GHz) reveal that the jets are not straight but display a few significant bends. The NW jet passes just outside, and to the west of, the three blobs in Region 4 and seems to come to an end on a level with the third blob. The SE jet first appears to pass the pronounced mass condensation next to the nucleus and then, a little bit further out, the protrusive structure in the NE flank of the dark filamentary system in Region 3. Since the jets are well collimated, they can by no means contribute to the formation of all the mammoth trunks scattered over the galaxy. Nevertheless, it may be of some interest to try finding out whether there exists some interaction between the jets and some of the dark features.

Geldzahler & Fomalont (1984) suggested that the bends of the jets could, at least partly, be caused by the deflection of the jets away from regions of gas and dust. In particular a sharp bend of the NW jet towards the west seems to coincide with the second blob in Region 4. The detailed HST, *ACS* image should here offer good opportunities to study a possible influence by the jets on the dark filaments. A close inspection of the image, however, reveals few or no visible signs of influence on the dusty gas along the paths of the jets. For instance, a number of fairly thin and weak filaments meandering eastwards and southwards from the pronounced mass condensation next to the nucleus show no signs of interaction with the SE jet although they are projected directly on the inner part of the jet. Likewise, there are a number of thin, loop-shaped filaments, seemingly connecting to the western side of the second blob, which show no traces of influence by the bending NW jet. Still, the jet appears to pass straight across the filaments. Only in the NE flank of the dark filamentary system in Region 3, there are a few thin filaments that are aligned with the SE jet. However, in the same region there also exist similar filamentary structures, which form substantial angles with the main direction of the jet. Thus, it is hard to find any visible sign of influence on the dark filaments, which can be positively coupled with the jets. On the whole, there are only a few dark features and no mammoth trunks along the projected paths of the jets.

### *5.3 Formation of ionized matter*

A vital task for the understanding of NGC 1316 is to find out how the ionized gas, discovered by Schweizer (1980), originated. Based on the agreement found between the velocities of the ionized gas and the CO gas and the disagreement between these velocities and the velocity pattern of the stars in the galaxy, Sage & Galletta (1993) argued that both gas components should have arisen in the same way, implying that they were both captured by the progenitor of NGC 1316. However, the same basic facts concerning the gas velocities and the velocity pattern of the stars could also be interpreted in a different way. The molecular gas component that is on the way of being captured, or has already been captured by NGC 1316, may give birth to the ionized component inside the galaxy. With this interpretation, only one gas component, viz. the molecular gas, entered the galaxy from outside.

When the small, gas-rich galaxy rushed into the progenitor of NGC 1316, the relative velocity must have been fairly large. CO observations show that velocities well exceeding 100 km $s^{-1}$ are involved in the motion of the filamentary, molecular gas. The motion of the gas, relative to a stationary plasma, must therefore be considered supersonic. At such high speeds, the kinetic energies of the particles are large enough to ionize atoms and molecules. In particular, Alfvén´s so called Critical Ionization Velocity (CIV) mechanism may here turn out to be important (Alfvén, 1942; 1981 and references therein).

An interesting possibility is that much of the ionized gas, observed by Schweizer, could result from the CIV mechanism, or a similar process. If so, mainly the front surfaces of the dark filaments should be ionized by the ambient plasma hitting the filaments. Besides the similarity of the velocity patterns, this hypothesis also finds some support in the fact that the large-scale distribution of the ionized gas is similar to the distribution of the CO gas contained in the dusty features.

So far, only one spectrogram has been published on the ionized gas in NGC 1316 (Schweizer, 1980). The spectrogram shows an intermittent O II emission along a slit crossing the central region of the galaxy, indicating that the ionized gas is not uniformly spread out in the galaxy. A more solid judgement of the above hypothesis would be possible if more detailed observations of the distribution of the ionized gas could be obtained. If it would turn out that the ionized gas exists in long bands or lanes mainly coinciding with the dark filamentary structures in NGC 1316, such observations would constitute strong evidence for the hypothesis. If, on the contrary, no positive correlation is found, a different explanation of the origin of the ionized gas has to be sought for.

In this connection, it may be of some interest to reflect on what would happen to the ions and electrons after they have been formed at the front surfaces of the dark filaments. In general, charged particles can move freely along magnetic field lines but have difficulties in moving perpendicular to them. Since after ionization, the ions and electrons must perform gyrating motions, they are likely to be trapped in various magnetic field configurations in the vicinity of the filaments. Gradually, probably after long time periods, the charged particles will diffuse out into the interstellar space of the galaxy. An important reason for the diffusion is that the particles are subject to drift motions, in particular the magnetic gradient drift and inertia drift. As a result of such processes, the ions and electrons released may form a tenuous, background plasma in the galaxy. It is tempting to identify this plasma with the ambient plasma considered in Section 4. If this interpretation is correct, the ambient plasma has probably not been formed through a single event. Instead, it is more likely that the plasma was gradually built up as the result of the infall of several gas-rich galaxies.

### *5.4 Mass distribution in the trunks*

In most of the mammoth trunks seen in NGC 1316, the head represents the darkest part and, hence, also the densest part. This is just what one should primarily expect from the double helix mechanism. The mass condensation in the head, together with the external force, compels the filament to be stretched into a V- or Y-shaped geometry. In addition to this, part of the filamentary matter in the tilted legs tends to slide along the magnetic field lines towards the head when the trunk is drawn out, a process making the mass condensation even more pronounced.

As regards the elephant trunks, there is also another effect going in the opposite direction. Due to the action of the central, luminous stars and the hot, expanding plasma inside the shell, the trunks are gradually ionized and eroded and finally shattered. Most exposed to this influence is the head of the trunks, facing the central OB stars. Hence, dusty matter at this site is expected to erode and fade first. For a Y-shaped filament, ionization of the gas at the top of the loop, connecting the two helical parts in the stem, may give an impression of a split head (CGK03). Actually, an invisible frame-work of closing magnetic field lines should still exist. A similar process may also account for the split heads found in a few of the mammoth trunks (Figure 6).

*5.5 Some odd structures*

The large, Y-shaped structure in the north-western quadrant, containing the western part of the prominent arc in Group 2, is an intriguing object (Figure 1). The structure certainly resembles the smaller, Y-shaped mammoth trunks in several respects but there are also obvious differences. For instance, the stem of the large, Y-shaped structure constitutes a direct continuation of the prominent arc while the stems of the smaller mammoth trunks stand out more or less perpendicularly to the dark filaments. Another difference regards the sizes of the objects. It is reasonable to believe that the large, Y-shaped structure and the mammoth trunks were not formed in exactly the same way. A possible interpretation of the large, Y-shaped structure is that the stem represents one of the more massive clouds in the colliding, small galaxy, e.g. a dense part of a former spiral arm, which has forced its way through the target galaxy while adjacent, less massive parts have got behindhand.

Two filamentary systems are to be found very close to the nucleus of NGC 1316 on the image in Figure 1. One of those is the massive system consisting of meandering filaments in Group 3 (Section 2). The head of the system appears to nearly touch the nucleus. Some similarity with the large, Y-shaped structure in Group 2 may here exist. The other system is constituted by the relatively weak, filamentary trunks just to the west of the nucleus and to the north of the three elongated clouds in Group 1. The trunks do not protrude perpendicularly from any filament like most of the other mammoth trunks do. Instead, their legs are bending southwards and seem to connect to the same web of thin filaments in the south as the three elongated clouds are connected to. The immediate impression is that the system of trunks appears to rush in towards the nucleus. It is quite possible that both of the systems, close to the nucleus, have lost most of their initial angular momentum and are therefore falling towards the nucleus. However, this supposition is presently hard to prove since we can only observe the projection of the clouds and have little knowledge of the velocities involved.

## 6 Conclusions

NGC 1316 is a giant, elliptical galaxy showing unambiguous signs of being a merger. The central region of the galaxy is sprinkled with a great number of dark and distinct features. A detailed study of these features has been performed using an HST, *ACS* image. In the first place, typical and recurrent structures have been searched for among all the features. From the study it has been found that:

1) The dark features are mainly constituted by filaments. Many of the filaments are nearly azimuthally oriented but other orientations exist as well.
2) Many of the filaments are collected into bundles or more complex systems.
3) There are a great number of relatively short, filamentary structures protruding inwards from the filaments, towards the central parts of the galaxy. Most of these structures have shapes resembling the letters V or Y. In some of the Y-shaped structures the stem of the Y can be resolved into intertwined filaments.
4) Many of the V- and Y-shaped structures are suggestive of elephant trunks, observed in H II regions. The V- and Y-shaped structures in NGC 1316 are, however, much larger than the elephant trunks and have therefore been termed mammoth trunks.
5) Although the mammoth trunks have a clear tendency of pointing inwards in the galaxy, there is a considerable scattering of their deviations from the radial direction. For a few trunks, the deviations are as large as ~90°.
6) A more regular pattern of the orientations of the mammoth trunks is obtained if the radial direction is abandoned as a norm and, if further, the dark features are divided into suitable sub-regions. Fairly large sub-regions exist, within which the orientations are well-ordered and in some cases even nearly parallel with each other.
7) There are also a number of mammoth trunks, which do not connect to any visible filament. Still, the structures seem to be arranged in fairly regular rows appearing to be extensions of dark, azimuthally oriented filaments.

Since the mammoth trunks are in many respects so similar to the elephant trunks, it is natural to assume that the mechanisms behind both of these kinds of object are also similar. In line with this, a new model of the mammoth trunks is proposed having several elements in common with a recent theory of elephant trunks. According to the model, the mammoth trunks were formed out of magnetized filaments acted upon by an external pressure force and internal inertia and magnetic forces. The magnetized filaments are identified with the long, dark filaments being present in NGC 1316 in great numbers. Most likely, the dark filaments are remnants from a recent collision and merging of the progenitor of NGC 1316 and a small, gas-rich galaxy. The external force, responsible for the adjustment of the mammoth trunks, is suggested to be due to the relative motion of the dark filaments and an ambient gaseous or plasma medium in the galaxy. In principle, this motion could be the result of either a galactic wind blowing out from the central parts of the galaxy, or an infall of the filaments through a comparatively stationary medium. The fact that the orientations of the mammoth trunks show a considerable scattering about the radial direction, gives some preference to the latter option. As a result of the external force, a non-uniform mass distribution, and an internal magnetic field the filaments are distorted into either a V-shape or a Y-shape. Both of these structures point inwards in the galaxy, thus forming the mammoth trunks. If the magnetic field inside the filament is only moderately twisted, or not twisted at all, the V-shaped type of trunk occurs while the Y-shaped type of trunk, containing a double helix in its stem, develops if the filamentary magnetic field is twisted beyond a certain critical limit. In this scenario, magnetic fields are of decisive importance for the sculpting of the mammoth trunks with all their intriguing shapes.

**Acknowledgements.** I wish to thank Anna Brink for kind assistance with the figures. My gratitude also goes to NASA, ESA, and The Hubble Heritage Team (STScI/AURA) for making the HST, *ACS* image of NGC 1316 available.